\documentclass{raa}
\usepackage{graphicx,times}
\usepackage{natbib}
\usepackage{amssymb,amsmath}
\bibpunct{(}{)}{;}{a}{}{,}

\usepackage[a4paper=true,dvipdfm=true,pagebackref=true]{hyperref}
\hypersetup{pdftitle = The title of my PDF, pdfauthor = My name, pdfsubject= The subject, pdfkeywords = keyword1 keyword2 keyword3}
\hypersetup{colorlinks = true, linkcolor = green, anchorcolor = red, citecolor = blue, filecolor = red, pagecolor = red, urlcolor = red}

\begin{document}

\title{Collapsing Index: A New Method to Identify Star-forming Cores Based on ALMA Images}
     \author{Nannan Yue\inst{1,2}, Yang Gao\inst{3}, Di Li\inst{1,2,4}, Liubin Pan\inst{3}
   }

 	\institute{ National Astronomical Observatories, CAS, Beijing 100012, People's Republic of China \email{gaoyang25@mail.sysu.edu.cn (Y.G.); dili@nao.cas.cn (D.L.)}
		\and University of Chinese Academy of Sciences, Beijing 100049, People's Republic of China
		\and School of Physics and Astronomy, Sun Yat-Sen University, Zhuhai, Guangdong 519082, People's Republic of China
		\and NAOC-UKZN Computational Astrophysics Centre, University of KwaZulu-Natal, Durban 4000, South Africa
	}

\abstract
{
Stars form through the gravitational collapse of molecular cloud cores.
Before collapsing, the cores are supported by thermal pressure and  turbulent motions.
A question of critical importance for the understanding of star formation is how to observationally
  discern whether a core has already initiated gravitational collapse or is still in hydrostatic balance.
The canonical method to identify gravitational collapse is based on the observed density radial profile,
  which would change from a Bonnor-Ebert type toward  power laws as the core collapses.
In practice, due to the projection effect, the resolution limit, and other caveats,
  it has been difficult to directly reveal the dynamical status of cores,
  particularly in massive star forming regions.
We here propose a novel, straight-forward diagnostic, namely, the collapsing index (CI),
  which can be modeled and calculated based on the radial profile of the line width of dense gas.
A meaningful measurement of CI requires spatially and spectrally resolved images
  of optically thin and chemically stable dense gas tracers.
ALMA observations are making such data sets increasingly available for massive star forming regions.
Applying our method to one of the deepest dense-gas spectral images ever taken toward such a region,
  namely, the Orion molecular cloud, we detect the dynamical status of selected cores therein.
  We observationally distinguished a collapsing core in a massive star forming region from a hydrostatical one. Our approach would help significantly improve our understanding
  of the interaction between gravity and turbulence within molecular cloud cores in the process of star formation.
  }

\keywords{stars: formation - ISM: molecules - ISM: kinematics and dynamics - turbulence}

\titlerunning{Collapsing Index: A New Method to Identify Star-forming Cores Based on ALMA Images}
\authorrunning{Yue N et al.}
\maketitle

\section{Introduction}

The dynamical evolution of a prestellar core is controlled by the interplay of turbulence, magnetic fields and gravity in the core, which is crucial for understanding the early phase of star formation \citep{maclow2004,Stahler2005,mckee2007}.
The initiation of the gravitational collapse of a prestellar core is of
particular interest, as it marks the transition toward a protostellar core
\citep{chandrasekhar1951,wardthompson2002,jouni2014,kritsuk2013,gao2015}.
 An important question is then how to observationally diagnose whether a prestellar core is collapsing or not. The dynamical status of a core may be identified based on theoretical predictions for the density and velocity structures
\citep{shu1977,foster1993,wardthompson1999}. Theoretical models make different predictions for the radial density profiles at different evolution stages of prestellar cores, and thus observations of the column density profile provide a tool to detect the dynamical status of the cores
\citep{alves2001,vazquez2001}.
This has motivated extensive observational studies of the column density, using far-infrared and submillimeter imaging, as well as dust extinction maps
\citep{evans2001,langer2001,alves2001}.
 However, due to the projection
effect and the resolution limit, the observed column density profile can hardly distinguish different theoretical models unambiguously,
except for a few special cases, where the Bonnor-Ebert type density profiles were observed through dust extinctions
\citep{alves2001}.

The gravitational collapse of a dense molecular cloud core may also
be detected using the velocity structures revealed by molecular lines. For example, the presence of a gravitational collapse velocity would cause the asymmetry of optically thick lines, 
leading to `blue profiles' \citep[][]{zhou1993,zhang1998,evans2002}. In this paper,
we develop a new method to identify the dynamical status of a
prestellar core using the radial distribution of the line width of optically thin molecular tracers. The width of molecular lines has contributions from both turbulent motions
\citep{larson1981,myers1983} and the gravitational infall velocity if it exists. We will show that the contributions from turbulence and the gravitational collapse to the line width are expected to have distinct spatial behaviors. In particular, 
gravitational infall motion would give rise to a rapid increase of line width toward the core center, whereas the contribution from turbulence remains almost invariant with decreasing radius. Therefore, the observed spatial distribution of line profiles may be used to determine the relative importance of turbulence and gravitational infall. Observations toward isolated cores using single dish radio telescopes have already measured the spatial distributions of line profiles \citep[e.g.][]{tafalla2004,kirk2009}.
The generally larger distances of massive star forming regions and more complex nature of dense gas require even higher spatial resolution and greater sensitivity, which were brought to realities by ALMA. We will define a collapsing index (CI) to characterize the dynamical status of two OMC cores.
By selecting cores with approximately spherical symmetry and simple environments,
it is possible to distinguish the gravitational infall motion from turbulence, providing a new tool to measure the dynamical status of the cores.

\section{Theoretical expectation} \label{sect:s2}

\begin{figure}
\centerline{	\includegraphics[width=6cm]{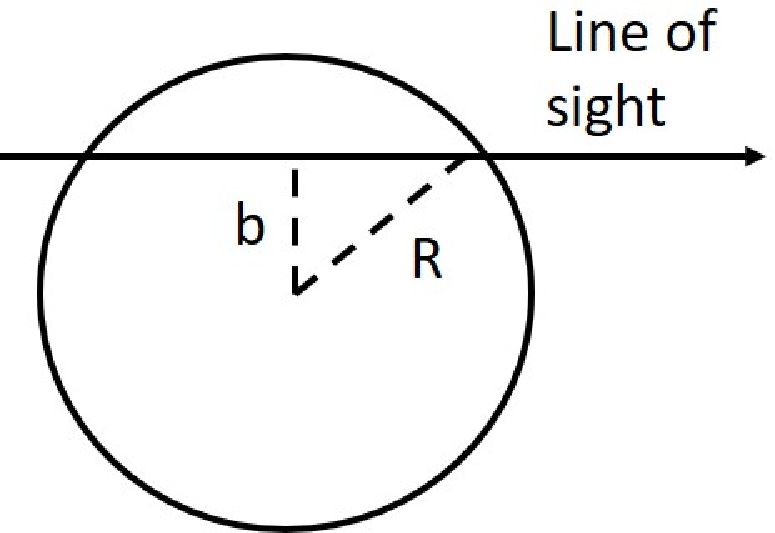}}
\centerline{     \includegraphics[width=8cm]{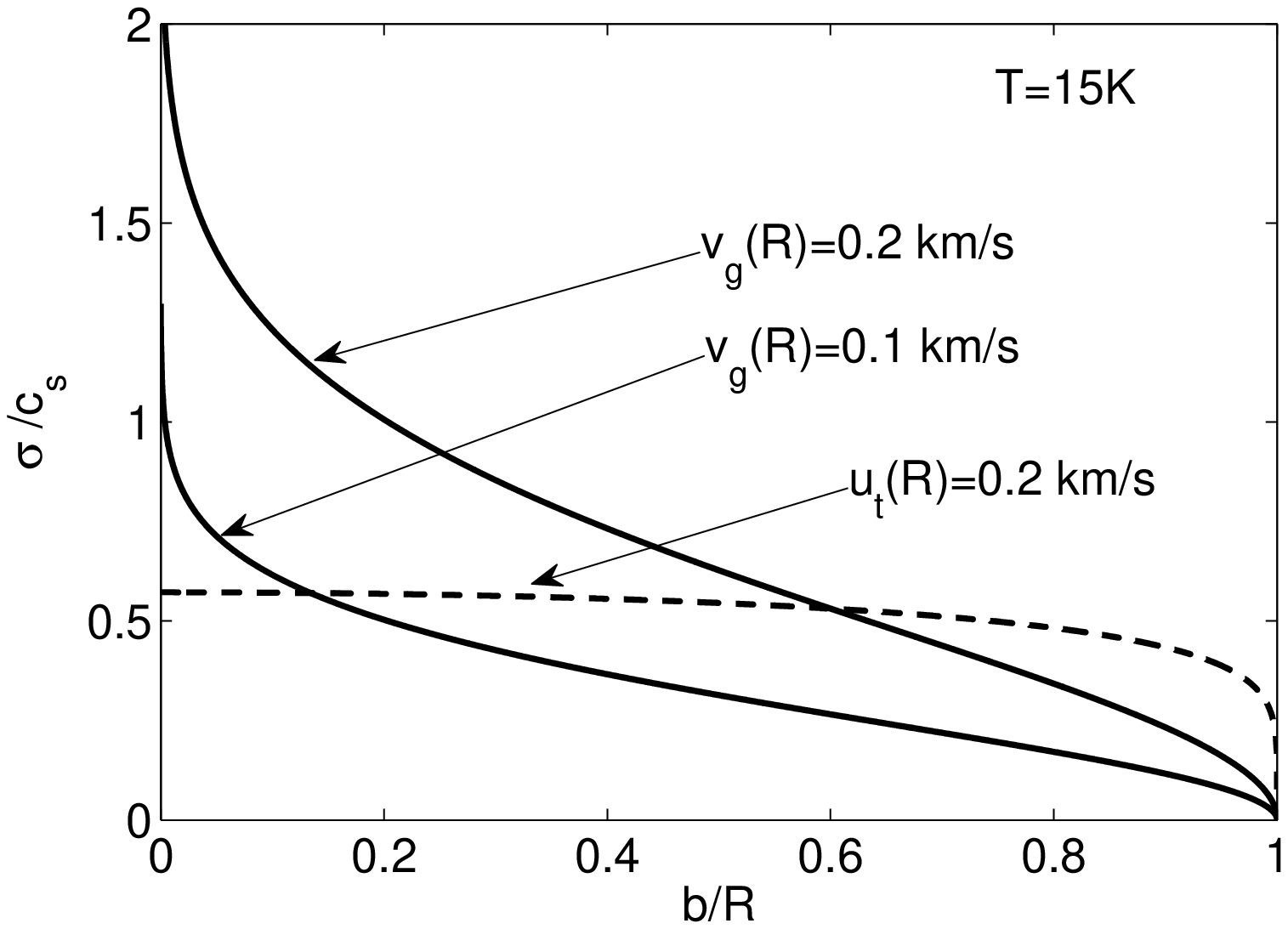}}
	\caption{\textbf{Upper}: line of sight with the impact parameter $b$ toward a cloud core of radius $R$.
\textbf{Lower}: theoretical expectation for the velocity dispersion (normalized to the speed of sound at
  15 K by assuming the ISM mean molecular weight being 2.33\,a.m.u.) as a function of the impact parameter.
  The solid and dashed lines show the contribution from gravitational infall in collapsing cores
    and turbulent velocity dispersion in non-collapsing cores, respectively.}
	\label{fig:theory}
\end{figure}

For a non-collapsing prestellar core, only turbulent and thermal motions contribute to the line width.
The contribution of turbulence to the width of an optically thin line is essentially the velocity dispersion
  over the length along the line of sight across the core (see the upper panel of Fig. \ref{fig:theory}).
To estimate the turbulent velocity dispersion,
  we make use of the Kolmogorov scaling law \citep{kolmogorov1941} for inertial-range scales,
 \begin{equation}
 u_{\rm t}(l) \propto (\bar \epsilon l)^{1/3},
 \label{equ:ut}
\end{equation}
  where $u_{\rm t}(l)$ is the amplitude of the velocity dispersion across a separation of $l$,
  and $\bar \epsilon$ is the average dissipation rate of turbulent kinetic energy.
Although originally developed for incompressible turbulence,
the Kolmogorov scaling also applies to subsonic or transonic turbulent flows \citep{padoan2004},
  which are more commonly seen in cloud cores.
Also implicit assumed in the Kolmogorov scaling is statistical homogeneity and isotropy,
  which reasonably applies to a non-collapsing core \citep{kritsuk2013}.
Applying Equation \ref{equ:ut} to the line of sight with impact parameter $b$ toward a core of radius $R$ (cf. Fig. \ref{fig:theory}),
  we find that the velocity dispersion $\sigma_t^2$ due to turbulence is given by,
  \begin{equation}
  \sigma_{\rm t}^2(b)=\frac{3}{11}u_{\rm t}^2(R)\left(\frac{2\sqrt{R^2-b^2}}{R}\right)^{\frac{2}{3}},
  \label{equ:utinter}
  \end{equation}
  where $u_{\rm t}(R)$ is the total amplitude of the turbulent velocity fluctuations at the core radius $R$,
  and the factor $\frac{3}{11}$ accounts for the fraction of velocity fluctuations along the longitudinal direction,
  i.e., along the line of sight.
  A detailed derivation of the above equation can be found in Appendix \ref{model}.
Equation \ref{equ:utinter} gives the expected contribution of turbulent motions to the line width as a function of the impact parameter $b$,
  which is shown as the dotted line in the lower panel of Fig. \ref{fig:theory}.
We find that if turbulent motions are supersonic, e,g., in a high mass star forming core, the velocity dispersion profile has a very similar trend to this case with subsonic turbulence. Therefore, the subsequent discussions based on subsonic turbulence would also qualitatively apply to the cores with supersonic turbulence.

If a core is undergoing gravitational collapse,
  both the infall motion and turbulence contribute to the observed line width.
To estimate the contribution of the collapse,
  we adopt the singular isothermal sphere (SIS) model of \citet{shu1977},
  for the collapse velocity,
  \begin{equation}
  v_{\rm g}(r)=v_{\rm g}(R)\left(\frac{r}{R}\right)^{-\frac{1}{2}},
  \label{equ:vg}
  \end{equation}
  which is essentially the free-fall velocity due to the central singular point mass.
  A power law density radial profile $\rho\propto r^{-\frac{3}{2}}$ is predicted in 
  this model,
  and an exact calculation of the contribution of the collapse velocity to the line 
 width needs to account for the density variations in the core.  
 However, for the first order approximation, 
 we neglect the effect of the density profile on the line width. 
Integrating  the square of the longitudinal component of the gravitational infall velocity (Equation \ref{equ:vg})
  along the line of sight gives a velocity dispersion, $\sigma_{\rm g}^2$,
  \begin{equation}
  \sigma_{\rm g}^2(b)=v_{\rm g}^2(R)\left(\frac{R}{\sqrt{R^2-b^2}}\rm{ln}\frac{R+\sqrt{R^2-b^2}}{b}-1\right).
  \label{equ:vginter}
  \end{equation}
As $b \to 0$, $\sigma_{\rm g}^2$ approaches infinity
  due to the singularity of the infall velocity in the self-similar collapse model.
For optically thin molecular lines, the equation predicts the line width from the
  collapse velocity as a function of the impact parameter, $b$.
As shown in Appendix \ref{model}, 
Equation \ref{equ:vginter}  well captures the behavior of the contribution 
of the infall velocity to the line width of a collapsing prestellar core, even though 
it ignores the density variations in the core.

Turbulence also contributes to the line width of a collapsing core.
From a simple dimensional analysis based on the turbulent energy source by the collapse velocity of the core \citep{robertson2012} and the similarity theory of Kolmogorov for turbulence, we speculate that the collapse-induced turbulent velocity has a similar radial dependence as $v_g$, i.e., $u_t \sim r^{-1/2}$. This similarity may be expected from the fact that the energy source of both the collapse velocity and the turbulent velocity is the gravitational energy. Our speculation is subject to tests by future numerical simulations of turbulence in a collapsing core.
Therefore, Equation \ref{equ:vginter} may be viewed as an approximate description for the radial dependence
  of the total velocity dispersion including the contributions from both the
  gravitational infall velocity and the collapse-induced turbulence.

As seen in the lower panel of Fig. \ref{fig:theory}, the velocity dispersions caused by turbulence in non-collapsing
  cores (dash line) and by gravitational infall in collapsing ones (solid lines) show very different radial
behaviors. Due to the faster infall velocity toward the center,
  the line broadening by the gravitational collapse increases rapidly with decreasing $b$.
In contrast, the turbulent velocity dispersion in the non-collapsing case is almost constant toward the center;
  it only decreases at the edge of the core,
  as the line of sight with smaller $b$ samples turbulence across a larger length scale.
Using the spatially resolved molecular line observations,
  the different trends of the velocity dispersion as a function of $b$ in collapsing and non-collapsing cores
  provide a unique tool to detect the dynamical status of the core.
The above analysis assumes that the molecular line is optically thin.
Our calculation finds that, as long as the optical depth is smaller than 2
  so that the molecular tracer may probe down to the center of the core,
  the expected spatial behavior of the line width remains qualitatively similar to that shown in Fig. \ref{fig:theory}.

In the ``inside-out" scenario of the SIS model,
  a non-collapsing outer envelope coexists with a collapsing inner core.
Therefore, the spatial distribution of the line width is expected to lie in between
Equation \ref{equ:utinter} and \ref{equ:vginter}.
We will assume that the turbulence velocity in the non-collapsing envelope may be described by Equation \ref{equ:utinter}
  based on statistically homogeneous turbulence.
Accounting for both the turbulence contribution from the envelope and the contribution of Equation \ref{equ:vginter}
  in the collapsing core region, the nonthermal line width can be approximately written as,
  \begin{equation}
  \begin{split}
  \sigma_{\rm NT}^2(b)=\frac{3}{11}u_{\rm t}^2(R)\left(\frac{2\sqrt{R^2-b^2}}{R}\right)^{\frac{2}{3}} \qquad \qquad \qquad   \\
  +v_{\rm g}^2(R)\left(\frac{R}{\sqrt{R^2-b^2}}\rm{ln}\frac{R+\sqrt{R^2-b^2}}{b}-1\right),
  \end{split}
  \label{equ:nth}
  \end{equation}
  where the parameters, $u_{\rm t}(R)$ and $v_{\rm g}(R)$,
  could be viewed as an indicator for the relative importance of turbulent and gravitational
  infall motions in a cloud core.

\section{Selected Orion Cores}

\begin{figure*}[ht!]
 \begin{center}
           \includegraphics[angle = 0, trim =0cm 0cm 0cm 0cm,width=15cm]{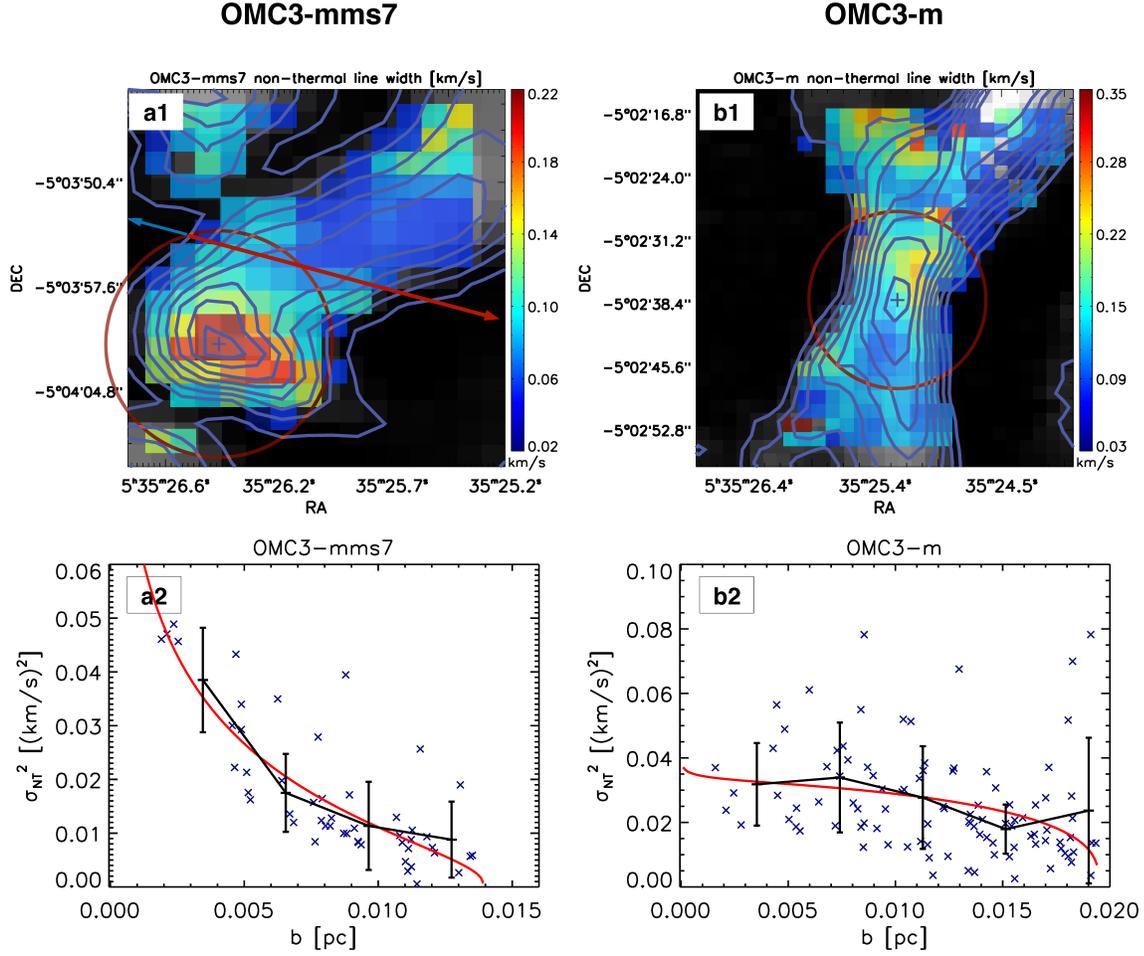}
       \end{center}
	\caption{Kinematics of two cores OMC3-mms7 and OMC3-m in Orion Molecular Cloud.
    The contour and the gray background in the upper row is the integrated intensity of ${\rm N_2H^+}$ (J=1-0),
    based on which the center (plus symbol) and edge (red circle) of each core are selected. 
   The centers of  OMC3-mms7 and OMC3-m are located at R. A. = $5^h 35^m 26^s.4$, DEC. =$-5^\circ 4'1''.2$  and R. A. = $5^h 35^m 25^s.3$, DEC. =$-5^\circ 2'36''.7$, respectively. 
   By fitting a Gaussian function to the average radial distribution of the ${\rm N_2H^+}$ (J=1-0) intensity, 
   the radii of OMC3-mms7 and OMC3-m are measured to be 0.014 pc and 0.020 pc, respectively (see Appendix \ref{radius}).
  The contour lines in \textbf{a1} start at 85 $\rm mJy\ beam^{-1}\ km\ s^{-1}$ with intervals of 200 $\rm mJy\ beam^{-1}\ km\ s^{-1}$, while the contour lines in \textbf{b1} start from 100 $\rm mJy\ beam^{-1}\ km\ s^{-1}$ with intervals of 120 $\rm mJy\ beam^{-1}\ km\ s^{-1}$.
    Color maps show the non-thermal line width dispersion. In \textbf{a1}, the blue and red arrows show the direction of the outflow originated from the center position at R. A. =$5^h 35^m 26^s.5$, DEC. =$-5^\circ 3'55''$ \citep{Takahashi2006,Takahashi2008}. It has a red lobe of 0.98 pc and a blue lobe of 0.15 pc with a position angle of 70$^\circ$ from north to northeast.
 In the bottom panels, the blue cross shows the non-thermal velocity dispersion from each observed pixel,
    while the black line plots the average velocity dispersion within each radial bin.
  The red curve is the best-fit for the radial profile using Equation \ref{equ:nth}.
  }
	\label{fig:omc23coresl}
\end{figure*}	

As the closest region containing massive young stellar clusters,
  the Orion Molecular Cloud (OMC) is one of the best laboratories to study star formation under the influence of massive proto-clusters.
The filament of OMC 2/3 is a relatively quiescent massive star forming region,
  containing dense cores of different dynamical statuses \citep{li2013}.
Combining ALMA and Nobeyama observations,
  we obtained a high spatial resolution map of the ${\rm N_2H^+}$ (J=1-0) line toward this region \citep{yue2020}.
The spatial resolution is $\sim$3", corresponding to 0.006 pc, smaller than the thermal Jeans scale $\sim 0.01$ pc;
  the rms noise is $\sim$12 mJy/beam, and the velocity resolution is 0.11 km/s.
For the cores selected for this study, our data are more than 2 times deeper compared to \citet{hacar2018} of 25 mJy/beam,
  thus provide better dynamic range.
The line fitting is carried out using a multi-Gaussian line fitting engine, SCOUSE \citep{henshaw2016}.
In our study, we choose two dense cores, OMC3-m and OMC3-mms7,
  that are roughly spherical and not significantly affected by the surrounding environment. 
  The selection of the cores and the measurement of their radii are further discussed in Appendix \ref{radius}.

Besides turbulence and possible gravitational motions,
  we also check the outflows and rotation motions which may exit in the two dense cores.
There is a bipolar outflow along the east-west direction associated with OMC3-mms7;
  but the center of this outflow is $\sim$ 10'' (0.02 pc) north,  and on the periphery of the selected dense core 
\citep{Takahashi2006}.
The existence of this outflow may inject energy into the core's turbulent motion, but is unlikely to cause systematic increase of the line width toward the center of OMC3-mms7, which we attribute to gravitational collapse.
There is no outflow around OMC3-m \citep{Takahashi2008}.
We also analyze the core rotation based on \cite{xu2020} and obtain the specific angular momentums (j = J/M), i.e., 0.0005 $\rm pc\ km/s$ for OMC3-m, and  0.003 $\rm pc\ km/s$ for OMC3-mms7, which correspond to less than 2\% of the gravitational potential energy for each core. The relatively insignificant rotational energy found here is consistent with the majority of the literatures \citep[e.g.,][]{Goldsmith1985,Tatematsu2016}. The rotational motions in these two Orion cores
are negligible in supporting against gravity and should not contribute to the observed trend of line width.

The observed molecular line width is mainly contributed by thermal, turbulent, and gravitational motions.
Subtracting the thermal contribution ($\sigma_{\rm T}$) from the observed line width ($\sigma_{\rm obs}$),
  we obtain the non-thermal line dispersion $\sigma_{\rm NT}$ \citep{myers1983},
  \begin{eqnarray}
  (\sigma_{\rm NT})^2 = (\sigma_{\rm obs})^2 - (\sigma_{\rm T})^2, \\
  \sigma_{\rm NT} = \sqrt{\frac{\Delta\upsilon^{2}_{\rm obs}}{\rm {8ln(2)}}-\frac{k_{\rm B}T_{\rm kin}}{m_{\rm obs}}}
  \label{equ:disp}.
  \end{eqnarray}
  where $\Delta\upsilon_{\rm obs}$ is the observed FWHM line width, $k_{\rm B}$ the Boltzmann constant,
  $T_{\rm kin}$ the gas kinetic temperature, and $m_{\rm obs}$ the mass of the observed molecule.
The kinematic information of the two selected cores is shown in the upper panels of Fig.\ \ref{fig:omc23coresl}.
The color maps represent the non-thermal dispersions from (\ref{equ:disp}).
The kinetic temperature, derived from JVLA NH$_3$ observations, varies only slightly within the cores \citep{li2013},
  and we adopt the average kinetic temperature of 15.7 K and 13.4 K for OMC3-mms7 and OMC3-m, respectively.

\section{Collapsing Index}

To observationally identify whether a prestellar core is collapsing,
  we define an empirical index, named the collapsing index (CI),
  to quantify the dynamical status of the core:
  \begin{equation}
  {\rm CI} = \frac{\sigma_{\rm in}^2-\sigma_{\rm out}^2}{\sigma_{\rm out}^2},
  \label{equ:gt}
  \end{equation}
  where $\sigma_{\rm in}^2=\left \langle \sigma_{\rm NT-inner}^2 \right \rangle$ is the average non-thermal line dispersion $\sigma_{\rm NT}^2(b)$ in the inner region with $b\le R/2$,
  while $\sigma_{\rm out}=\left \langle \sigma_{\rm NT-outer}^2 \right \rangle$ is the average over the outer region
  with $b> R/2$.
Based on the lower panel of Fig.\ \ref{fig:theory}, it is expected that a core undergoing collapse would have a larger CI.
Inserting Equation \ref{equ:utinter} and \ref{equ:vginter}
into the definition of the index gives CI = 0.3 and 5.3, respectively.
We adopt 0.3, the expected CI if only turbulence contributes to $\sigma_{\rm NT}$,
  as the critical value to determine whether a prestellar core is collapsing.
Clearly, if the measured CI in a core is much larger than 0.3,
  it indicates a significant contribution from the collapse velocity to the line width.

A calculation using the observed $\sigma_{\rm NT}$ for the selected Orion cores shows that CI = 1.90 and 0.32 for OMC3-mms7 and OMC3-m, respectively.
For OMC3-mms7, we find that CI is much larger than 0.3, suggesting that
  gravitational collapse has already initiated in this core.
On the other hand, the CI of OMC3-m is close to 0.3, indicating that its non-thermal line width is dominated by turbulence,
  and the core has not started to collapse.
We stress that it is the first time the dynamical status of prestellar cores of radius $\leq$0.02 pc is identified.
As a supportive clue of its dynamic status, the asymmetric profiles of the molecular lines $^{13}$CO (J=1-0) and C$^{18}$O (J=1-0) in the dense core region show signatures of possibly infall for OMC3-mms7.
On the other hand, for OMC3-m, the $^{13}$CO and C$^{18}$O lines are symmetric, likely indicating no global infall motions in it (see Appendix \ref{co}).

\begin{figure}
\centerline{	\includegraphics[width=12cm]{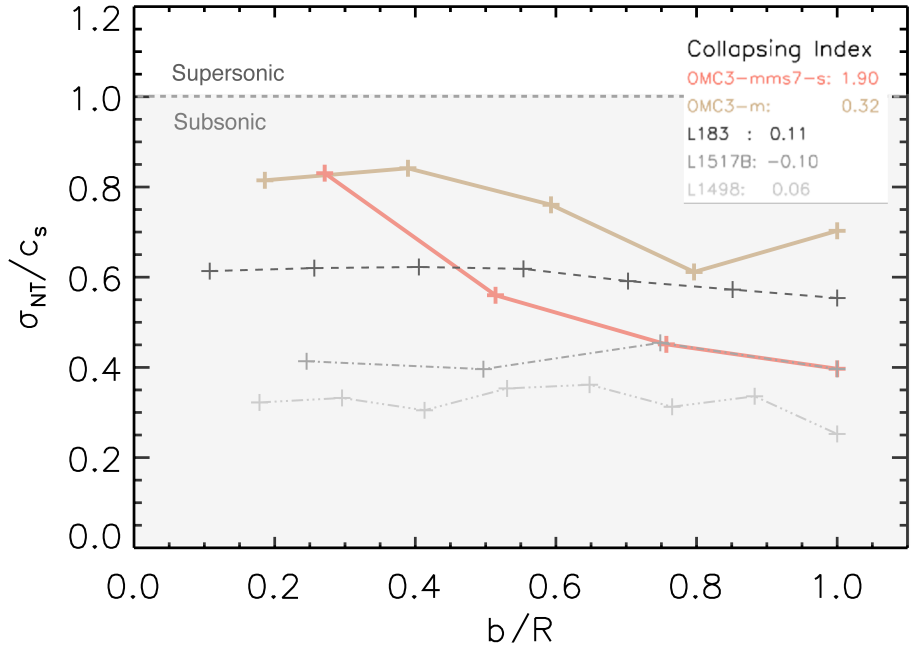} }
	\caption{The radial distribution of non-thermal velocity dispersion, normalized to the speed of sound, for 5 cores.
The Collapsing Index (CI) for each core is indicated in the legend.
Both Orion cores (in thick colored lines) and isolated cores (dashed thin lines) are
  in subsonic turbulence region (i.e.\ non-thermal line width smaller than thermal line width).
Only OMC3-mms7 have CI greater than 0.3, indicating collapse.}
	\label{fig:2panel_core}
\end{figure}

We also examined three isolated starless cores, i.e., the high Galactic latitude core L183,
  and two cores in the Taurus-Auriga cloud complex, L1498 and L1517B,
  with different physical environments from the OMC.
Due to their smaller distance ($\sim 150$ pc) and larger size ($R =0.04-0.06$ pc),
  these cores are spatially resolved by single dish telescopes.
For the three cores, we make use of the observation results from \citet{tafalla2004,tatematsu2016}
  for the spatial distribution of the ${\rm N_2H^+}$ (J=1-0) line width;
  and adopt the nearly uniform kinetic temperature of 7 K, 9.4 K and 10.7 K for L183, L1517B and L1498, respectively \citep{tafalla2004,kirk2009}.
The CI for all these isolated starless cores is found to be significantly smaller than 0.3,
  suggesting that they are not collapsing (cf. Fig.\ \ref{fig:2panel_core}).
 It is noted that L1498 seems to show signatures of infall from its CS (J=2-1) lines asymmetries \citep{caselli2002};
  and L183 is also thought to be contracting according to its possible rotation motion and spin up \citep{kirk2009}.
  However, as an effective tracer of infall motion, no HCO$^+$ (J=1-0) asymmetry is found in L1498 or L183,
    indicating them either being non-collapsing, or contracting at a very small amplitude of infall speed.
  For L1517B, the HCO$^+$ (J=1-0) lines show red asymmetry which indicates the existence of
    both envelope expansion and core collapse (EECC) motions \citep{tafalla2004,fu2011}.
    With the infall speed being small, the global EECC motion usually occurs at the very initial stage of core collapse \citep{gao2010}.
  Our CI method, as shown in Fig.\ \ref{fig:2panel_core}, does not identify these possible initial stage, small amplitude core contracting motion or EECC motion
    as core collapse.

Fig.\ \ref{fig:2panel_core} shows the radial profile of the non-thermal velocity dispersion,
  normalized to the sound speed $c_{\rm s} \equiv \sqrt{\frac{k_{\rm B}T_{\rm kin}}{m_{\rm mol}}}$
  with $m_{\rm mol}$ set to 2.33\,a.m.u., for all the 5 cores selected in our study.
Note that, the non-collapsing core, OMC3-m, has a velocity dispersion larger than the three isolated cores,
likely due to the more violent environment in the OMC.
The velocity dispersions in all the cores are subsonic,
  which validates our theoretical treatment based on the scaling behavior of incompressible or weakly compressible turbulence.

\begin{table}
\small
\caption{Strength of turbulent and gravitational motions in the selected OMC cores by model fitting.
} \label{tab:fit}
\small
\begin{center}
\begin{tabular}{cccc}
\hline
\hline
				  & $u_{\rm t}(R) [km/s]$ & $v_{\rm g}(R) [km/s]$	& $\chi^2$ \\ \hline
 OMC3-mms7 & 0.12 &  0.16 & 0.5 \\ \hline
 OMC3-m &	0.27 &	 0.03 &	2.6 \\ \hline
\end{tabular}
\end{center}
\end{table}

The importance of the collapse velocity relative to turbulent motions can also be estimated by fitting
  the observed radial profile of the line width with our model, Equation \ref{equ:nth}.
Using the non-linear least-square fitting algorithm, mpfit \citep[][]{markwardt2009},
  we tune the two parameters, $u_{\rm t}(R)$ and $v_{\rm g}(R)$,
  in the equation to yield the best fits,
  which are shown as red lines in the bottom panels, a2 and b2 of Fig.\ \ref{fig:omc23coresl}.
The best-fit parameters are listed in Tab.\ \ref{tab:fit}.
For OMC3-mms7, the gravitational collapse velocity is slightly larger than the turbulent velocity even at the edge of the core,
  suggesting that it is already undergoing gravitational collapse.
In OMC3-m, $v_{\rm g}(R)$ is negligible, and the non-thermal velocity dispersion is dominated by turbulence,
  implying that the core is still in hydrostatic equilibrium.
  
  In our analysis, the motion of rotation is not included. If rotation is considered, at least three more parameters, i.e. the scaling and radial distribution of the rotation velocity, and the inclination angle of the rotation axis have to be added in the model. The increase of complication will smear the different features of gravity and turbulence found in this paper. The case is similar for bipolar outflows. So in the current study we prefer to aim at the cores where rotation is not significant and bipolar outflows affect little.

The morphology deviation of our selected cores from ideally spherical symmetry is not significant. We adopt the recipe in \cite{li2013} and \cite{li2002}, which derived an analytical relation between the intrinsic ellipticity and the average of its random projections. The observed axis ratio ($f_{obs}$) between short and long axes of the cores in our sample are 0.82 for OMC3-mms7 and 0.78 for OMC3-m, which correspond to a plausible intrinsic axis ratio ($f$) of 0.75 and 0.7. The resulting increase of gravitational potential energy scales with $\beta = arcsin(\sqrt{1-f^2})/ (\sqrt{1-f^2})$. The correction factor due to ellipsoid considerations are 1.09 for OMC3-mms7 and 1.11 for OMC3-m, respectively. The virial parameters (gravitational potential / kinetic energy) increase around $9\%$ and $11\%$. They are not severely affected by the deviation from spherical symmetry. From the observational point of view, we consider the selected cores in this paper as spherical symmetric ones.

\section{Conclusion}

Based on spatially resolved line width maps of molecular cloud cores,
  an empirical diagnostic, named the Collapsing Index (CI),
  is defined to quantify the importance of the gravitational collapse velocity relative to turbulent motions.
Spatially and spectrally resolved images from ALMA observations of ${\rm N_2H^+}$ (J=1-0) lines,
  the best tracer to probe deep into the dense cores \citep{caselli2002}, make it possible to evaluate the CI,
  allowing us to tell whether a prestellar core is undergoing gravitational collapse.
Other dense core tracers such as ${\rm NH_3}$ and ${\rm H_2D^+}$ may also be used in the CI analysis.
In the 5 cores studied, we found that all isolated low mass cores have CI smaller than 0.3,
indicating that they are in hydrostatical balance;
  while one OMC core has a CI much larger than 0.3, indicating it is collapsing.
The turbulent motions in all the 5 cores are subsonic.

We have restricted our study to a selection of prestellar cores with spherical morphology that are not significantly affected by rotation or bipolar outflows.  
Our approach may be generalized to account for the effects of rotation and bipolar outflows, as well as the more complicated morphology of the cores, which, however, is out of the scope of the current study.
The new approach developed in this work will help us clarify  
  the initiation of gravitational collapse in molecular cloud cores,
  and is thus expected to significantly advance our understanding of star formation in the prestellar stage.

\begin{acknowledgements}
This work is supported by the National Natural Science Foundation of China grant No. 11988101, and by the startup fund from Sun Yat-Sen University. LP acknowledges financial support from NSFC under grant No. 11973098.
Dr. Tie Liu (SHAO), Dr. Lei Qian (NAOC), Dr. Zhiyuan Ren (NAOC), Dr. Zhichen Pan (NAOC) and Ms. Xuefang Xu (NAOC)  are acknowledged for helpful discussions and suggestions. This paper makes use of the following ALMA data: ADS/JAO.ALMA\#2013.1.00662.S. ALMA is a partnership of ESO (representing its member states), NSF (USA) and NINS (Japan), together with NRC (Canada), MOST and ASIAA (Taiwan), and KASI (Republic of Korea), in cooperation with the Republic of Chile. The Joint ALMA Observatory is operated by ESO, AUI/NRAO and NAOJ.
We thank Dr. Shuo Kong and CARMA-NRO Orion team to provide the $^{13}$CO and C$^{18}$O datacubes in this region. 
\end{acknowledgements}

\bibliographystyle{aa}
\bibliography{myrefs}

\begin{thebibliography}{}

\bibitem[Alves et al.\ (2001)]{alves2001}
    Alves, J., Lada, C. J. \& Lada, E. A.\ 2001, Nature, 406, 159

\bibitem[Caselli et al.\ (2002)]{caselli2002}
    Caselli, P., Benson, P.~J., Myers, P.~C. \& Tafalla, M.\ 2002, \apj, 572, 238

\bibitem[Chandrasekhar (1951)]{chandrasekhar1951}
    Chandrasekhar, S. 1951b, P. Roy. Soc. Lond. A Mat., 210, 26

\bibitem[Evans et al.\ (2001)]{evans2001}
    Evans N. J., II, Rawlings, J. M. C., Shirley, Y. L. \& Mundy, L. G.\ 2001, \apj, 557, 193

\bibitem[Evans (2002)]{evans2002}
    Evans N. J., II, 2003, in Curry C. L., Fich M., eds, Proc. Conf., Chemistry as a Diagnostic of Star Formation.
    NRC Research Press, Ottawa, Canada, p. 157

\bibitem[Foster \& Chevalier (1993)]{foster1993}
    Foster, P. N. \& Chevalier R. A.\ 1993, ApJ, 416, 303

\bibitem[Fu et al.\ (2011)]{fu2011}
    Fu, T., Gao, Y. \& Lou, Y.-Q. 2011, ApJ, 741, 113

\bibitem[Gao \& Lou (2010)]{gao2010}
    Gao, Y. \& Lou, Y.-Q. 2010, MNRAS, 403, 1919

\bibitem[Gao et al.\ (2015)]{gao2015}
    Gao, Y., Xu, H. \& Law, C. K. 2015, ApJ, 799, 227

\bibitem[Goldsmith \& Arquilla(1985)]{Goldsmith1985}
	Goldsmith, P.~F., \& Arquilla, R.\ 1985, Protostars and Planets II, 137

\bibitem[Hacar et al.\ (2018)]{hacar2018}
    Hacar, A., Tafalla, M., Forbrich, J., et al. 2018, A\&A, 610, A77

\bibitem[Henshaw et al.\ (2016)]{henshaw2016}
  Henshaw, J. D., Longmore, S. N., Kruijssen, J. M. D. et al. 2016,
  Astrophysics Source Code Library (ASCL), 1601, 003

\bibitem[Jouni et al.\ (2014)]{jouni2014}
 K. Jouni, F. Federrath, T. Henning, 2014, \emph{Science}, 344, 183 

\bibitem[Kirk et al.\ (2009)]{kirk2009}
  Kirk, J. M., Crutcher, M. \& Ward-Thompson, D. 2009, ApJ, 701, 1044

\bibitem[Kolmogorov (1941)]{kolmogorov1941}
    Kolmogorov, A. N. 1941, DoSSR, 30, 301

\bibitem[Kong et al.(2018)]{kong2018}
 Kong, S., Arce, H.~G., Feddersen, J.~R., et al.\ 2018, \apjs, 236, 25


\bibitem[Kritsuk et al.\ (2013)]{kritsuk2013}
    Kritsuk, A. G., Lee, C. T. \& Norman L. M. 2013, MNRAS, 436, 3247

\bibitem[Landau \& Lifshitz (1987)]{landau1987}
   Landau, L. D. \& Lifshitz, E. M. 1987,Fluid Mechanics (2nd eds.), Elsevier, New York

\bibitem[Langer \& Willacy (2001)]{langer2001}
   Langer, W. D. \& Willacy, K. 2001, ApJ, 557, 714

\bibitem[Larson (1981)]{larson1981}
    Larson, R. B. 1981, MNRAS, 194, 809
    
\bibitem[Li(2002)]{li2002} Li, D.\ 2002, Ph.D. Thesis    

\bibitem[Li et al.\ (2013)]{li2013}
    Li, D., Kauffmann, J., Zhang, Q., \& Chen, W.\ 2013, \apjl, 768, L5 
    
\bibitem[Mac Low \& Klessen (2004)]{maclow2004}
    Mac Low, M.-M. \& Klessen, R. S. 2004, Rev. Mod. Phys., 76, 125

\bibitem[McKee \& Ostriker (2007)]{mckee2007}
    McKee, C. F. \& Ostriker, E. C. 2007, ARA\&A, 45, 565

\bibitem[Markwardt (2009)]{markwardt2009}
Markwardt C. B. 2009, in Bohlender D. A., Durand D., Dowler P.,
eds, Astronomical Society of the Pacific Conference Series Vol. 411, Astronomical Data Analysis Software and Systems XVIII. p. 251 (arXiv:0902.2850)

\bibitem[Myers (1983)]{myers1983}
    Myers, P.~C.\ 1983, \apj, 270, 105

\bibitem[Padoan et al.\ (2004)]{padoan2004}
  Padoan, P., Jimenez, R., Nordlund, {\AA}. \& Boldyrev, S. 2004, PRL, 92, 191102

\bibitem[Robertson \& Goldreich (2012)]{robertson2012}
   Robertson, B. \& Goldreich, P. 2012, ApJ, 750, 31

\bibitem[Stahler \& Palla(2005)]{Stahler2005}
Stahler, S.~W., \& Palla, F.\ 2005, The Formation of Stars

\bibitem[Shu (1977)]{shu1977}
   Shu, F. H. 1977, ApJ, 214, 488

\bibitem[Tafalla et al.\ (2004)]{tafalla2004}
    Tafalla, M., Myers, P.~C., Caselli, P., \& Walmsley, C.~M.\ 2004, \aap, 416, 191

\bibitem[Takahashi et al.(2006)]{Takahashi2006}
    Takahashi, S., Saito, M., Takakuwa, S., et al.\ 2006, \apj, 651, 933

\bibitem[Takahashi et al.(2008)]{Takahashi2008}
Takahashi, S., Saito, M., Ohashi, N., et al.\ 2008, \apj, 688, 344

\bibitem[Tatematsu et al.\ (2016)]{tatematsu2016}
    Tatematsu, K., Liu, T., Ohashi, S., et al. 2016, ApJS, 228, 2

\bibitem[Tatematsu et al.(2016)]{Tatematsu2016} Tatematsu, K., Ohashi, S., Sanhueza, P., et al.\ 2016, \pasj, 68, 24

\bibitem[Vazquez-Semadeni \& Garcia (2001)]{vazquez2001}
     Vazquez-Semadeni, E. \& Garcia, N. 2001, ApJ, 557, 727

\bibitem[Ward-Thompson et al.\ (1999)]{wardthompson1999}
    Ward-Thompson, D., Motte, F., Andr\'{e}, P. 1999, MNRAS, 305, 143

\bibitem[Ward-Thompson\ (2002)]{wardthompson2002}
Ward-Thompson, D., 2002, \emph{Science}, 295, 76

\bibitem[Xu et al.(2020)]{xu2020} Xu, X., Li, D., Dai, Y.~S., et al.\ 2020, arXiv e-prints, arXiv:2004.14643

\bibitem[Yue et al.\ (2020)]{yue2020}
  Yue, N., Li, D., Zhang, Q. et al. 2020, accepted by RAA, arXiv:2006.04168

\bibitem[Zhang et al.\ (1998)]{zhang1998}
   Zhang, Q., Ho, P. T. P. \& Ohashi, N. 1998, ApJ, 494, 636

\bibitem[Zhou et al.\ (1993)]{zhou1993}
    Zhou, S., Evans, N. J. II., Kompe, C. \& Walmsley, C. M. 1993, ApJ, 404, 232

\end{thebibliography}

\appendix
\section{Derivations of equations in Section 2 and the effects of the core density profile}{\label{model}}

\begin{figure}[ht!]
 \begin{center}
           \includegraphics[angle = 0, trim =0cm 0cm 0cm 0cm,width=6cm]{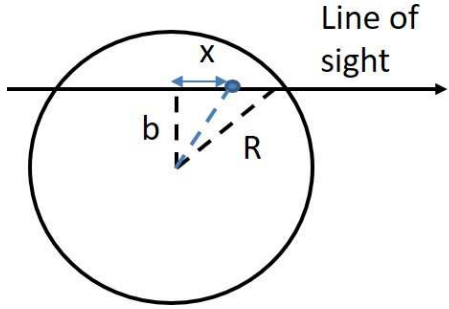}
       \end{center}
	\caption{A line of sight with an impact parameter $b$ towards a spherical core (the circle) of radius $R$.   The contribution 
	of  turbulent motions to the line width along this line of sight is estimated as the turbulent velocity dispersion 
	 across the length, $2\sqrt{R^2-b^2}$, of the chord intercepted (Equation \ref{equ:app_turb}). 
         The contribution to the line width  due to the gravitational infall is calculated by an integration 
         along the chord, using the velocity profile of the singular isothermal sphere  model (Equation \ref{equ:app_nodensity} and Equation \ref{equ:app_density}).}
	\label{fig:app_int}
\end{figure}	

As illustrated in Fig.\ \ref{fig:app_int}, the intersection of the line of sight with an impact parameter $b$ with a spherical core of 
radius $R$ makes a chord of length, $2\sqrt{R^2-b^2}$. Applying the Kolmogorov scaling law (Equation \ref{equ:ut}) 
to this length scale, we find that the turbulent velocity dispersion 
$u_t^2(b)$ is,
\begin{equation}
   u_t^2(b)=u_{\rm t}^2(R)\left(\frac{2\sqrt{R^2-b^2}}{R}\right)^{\frac{2}{3}},
   \label{equ:app_turb}
\end{equation}
with $u_{\rm t}^2(R)$ the turbulent velocity dispersion at the scale of the core radius $R$. 
The velocity dispersion is composed of the longitudinal and transverse structure functions,
$u_{\rm ll}^2$ and $u_{\rm nn}^2$ respectively, by
 \begin{equation}
    u_{\rm t}^2 = u_{\rm ll}^2 + 2u_{\rm nn}^2.
  \end{equation}
For incompressible or weakly compressible turbulence, 
we have \citep{landau1987}
  \begin{equation}
  u_{\rm nn}^2=\frac{4}{3}u_{\rm ll}^2.
  \end{equation}
for scales in the inertial range. The observed velocity dispersion along a line sight with impact parameter 
$b$ corresponds to the longitudinal component $u_{\rm ll}^2$ of the structure function at a separation 
of $2\sqrt{R^2-b^2}$ (cf. Fig.\ \ref{fig:app_int}). We thus have, 
 \begin{equation}
  \sigma_{\rm t}^2(b)=\frac{3}{11}u_{\rm t}^2(R)\left(\frac{2\sqrt{R^2-b^2}}{R}\right)^{\frac{2}{3}},
  \end{equation}
which is Equation \ref{equ:utinter} in Section \ref{sect:s2}. 

\begin{figure}[ht!]
 \begin{center}
           \includegraphics[angle = 0, trim =0cm 0cm 0cm 0cm,width=10cm]{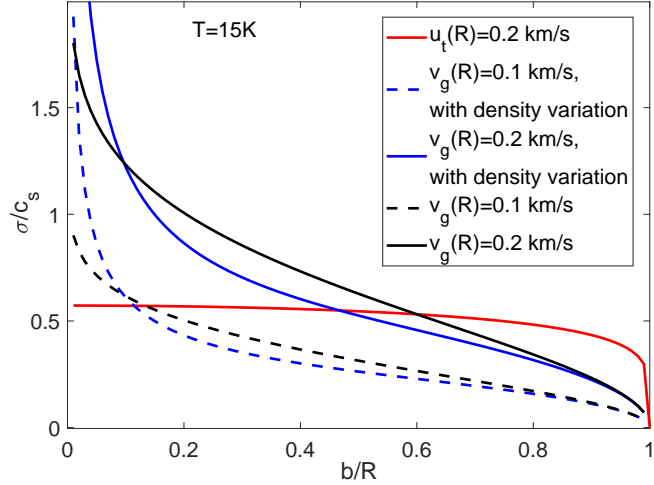}
       \end{center}
	\caption{The contributions of  turbulence (red line) and gravitational infall (blue and black lines) to the line width. 
	The blue and black lines are based on Equation \ref{equ:app_density} and Equation \ref{equ:vginter} in the tex 
	with and without density weighting, respectively.}
	\label{fig:density_variation}
\end{figure}	

Based on the radial velocity profile of the singular isothermal sphere (Equation \ref{equ:vg}), we integrate the square of the longitudinal 
component of the collapse velocity along the intercepted chord, i.e.,
 \begin{equation}
 \sigma_{\rm g}^2(b)=\frac{1}{2\sqrt{R^2-b^2}}\int_{-\sqrt{R^2-b^2}}^{\sqrt{R^2-b^2}}v_{\rm g}^2(R)\left(\frac{\sqrt{b^2+x^2}}{R}\right)^{-1}\frac{x^2}{x^2+b^2}dx,
 \label{equ:app_nodensity}
  \end{equation}
 where the integration variable $x$ is illustrated in Fig.\ \ref{fig:app_int}.
The integral in the above equation can be calculated analytically, and we find that 
 \begin{equation}
 \sigma_{\rm g}^2(b)=v_{\rm g}^2(R)\left(\frac{R}{\sqrt{R^2-b^2}}\rm{ln}\frac{R+\sqrt{R^2-b^2}}{b}-1\right),
 \end{equation}
 which is Equation \ref{equ:vginter} in the text.

An accurate estimate for the contribution of the collapse velocity to the line width needs to account for 
the radial density variations in the core along the line of sight.  
Using the density radial profile, $\rho\propto r^{-\frac{3}{2}}$,  in the SIS model of \citet{shu1977} and 
applying a density weighting factor of $\rho^2$, the contribution to the line width by the gravitational infall is given by, 
  \begin{equation}
  \sigma_{\rm g(\rho)}^2(b)=\frac{\int_{-\sqrt{R^2-b^2}}^{\sqrt{R^2-b^2}}v_{\rm g}^2(R)\left(\frac{\sqrt{b^2+x^2}}{R}\right)^{-1}\frac{x^2}{x^2+b^2}\left(\frac{\sqrt{b^2+x^2}}{R}\right)^{-3}dx} {\int_{-\sqrt{R^2-b^2}}^{\sqrt{R^2-b^2}}\left(\frac{\sqrt{b^2+x^2}}{R}\right)^{-3}dx}.
  \label{equ:app_density}
  \end{equation}
The numerator in this equation cannot be integrated analytically.  The numerical evaluation of the density weighted velocity dispersion 
(Equation \ref{equ:app_density}) by the gravitational collapse is shown as the blue curves in Fig.\ \ref{fig:density_variation}.
The black curves show the prediction of  Equation \ref{equ:vginter} in the text, which ignores the density variations in the 
core.  As seen in Fig.\ \ref{fig:density_variation}, the behaviors of  the blue and black lines with and 
without density weighting are qualitatively very similar, suggesting that, as a first order approximation, 
Equation \ref{equ:vginter}  in the text well captures the key features of the contribution of the infall to the line width of a collapsing 
prestellar core. 

\section{Measurement of the core Radius}{\label{radius}}

\begin{figure*}[ht!]
 \begin{center}
    \includegraphics[angle = 0, trim =0cm 0cm 0cm 0cm,width=15cm]{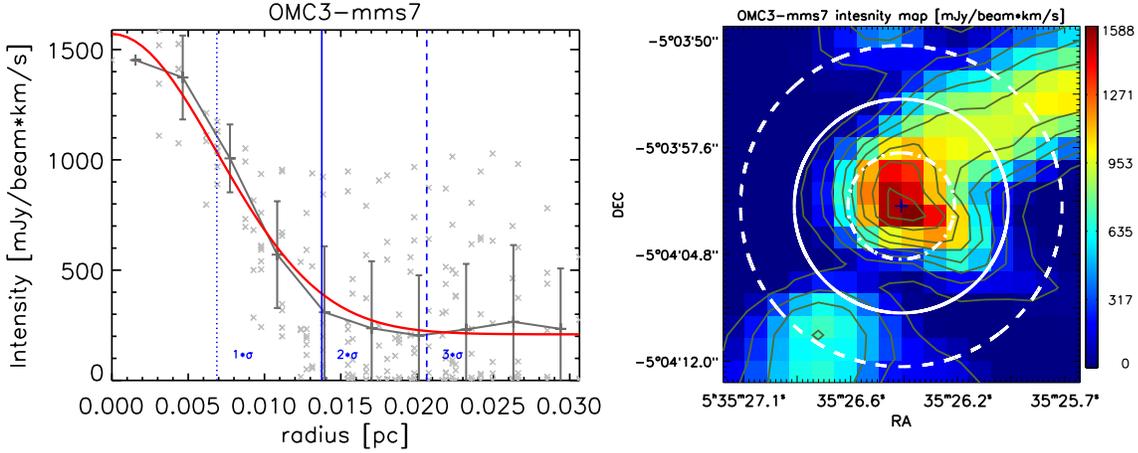}
  \end{center}
	\caption{\textbf{Left:} The radial distribution of the $\rm N_2H^+$ intensity for the core OMC3-mms7. The gray dots are the observed $\rm N_2H^+$  intensity at each pixel. The black line is the average intensity within the bin size of $\rm \sim 0.003$ pc, while the error bar indicates the rms 
	of the intensity variations in each bin. 
	The mean noise (rms) of the integrated map in this region is $\sim$ 12 $\rm mJy\ beam^{-1}\ km\ s^{-1}$. The red line is the gaussian function with a standard deviation of $\sigma = \sim$ 0.007 pc, which best fits the black line.
	The blue lines represent radii of 1, 2 and 3$\sigma$ to the core center. The blue solid line is  adopted as the core radius ($R=$0.014 pc)  in the main text. \textbf{Right:} The $\rm N_2H^+$ intensity map of OMC3-mms7. The contour starts from 85 $\rm mJy\ beam^{-1}\ km\ s^{-1}$ with a step of 200 $\rm mJy\ beam^{-1}\ km\ s^{-1}$. 
	The white lines correspond to radii of 1, 2 and 3$\sigma$.
}
	\label{fig:r_omc3_mms7}
\end{figure*}

\begin{figure*}[ht!]
 \begin{center}
           \includegraphics[angle = 0, trim =0cm 0cm 0cm 0cm,width=15cm]{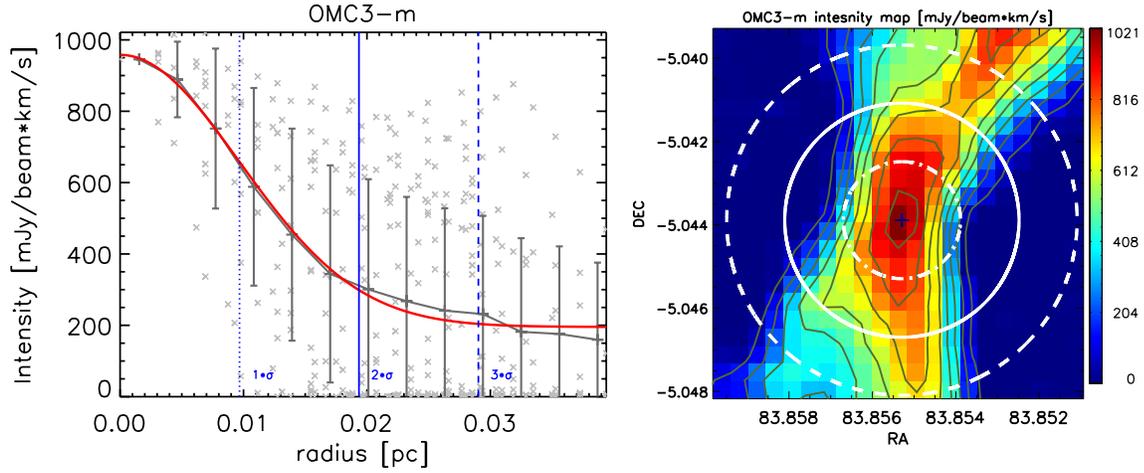}
       \end{center}
	\caption{Same as Figure \ref{fig:r_omc3_mms7}, but for OMC3-m. \textbf{Left:}  The radial distribution of the average $\rm N_2H^+$ intensity (black line)  
 	 is  fit by a Gaussian function with a standard deviation of $\sigma =\sim$0.01 pc.
	 The blue lines are located  at distances of 1, 2 and 3$\sigma$ from the center. The blue solid line is adopted as  the  core radius (0.02 pc) 
	 in the main text. \textbf{Right:} The $\rm N_2H^+$ intensity map of OMC3-m. The contour starts from 100 $\rm mJy\ beam^{-1}\ km\ s^{-1}$ with a step of 120 $\rm mJy\ beam^{-1}\ km\ s^{-1}$.  
	 The white lines are locations of 1, 2 and 3$\sigma$ from the core center.}
	\label{fig:r_omc3_m}
\end{figure*}

In our study, we selected two starless Orion cores based on the integrated intensity 
of $\rm N_2H^+$ emission lines. To avoid complexities, the two selected Orion cores are relatively 
isolated and have approximately spherical morphology. The high-resolution observational data allow a study of the 
radial profile of the molecular emission lines. 
We treat the selected cores as spherically symmetric objects and identify the position 
where the molecular line emission peaks as the center. 
We find that the intensity of the emission line as a function of the 
distance to the core center may be fit by a Gaussian function (see the red lines in Fig.\ \ref{fig:r_omc3_mms7} and \ref{fig:r_omc3_m}). 
In the text, we take the core radius, $R$, to be $2 \sigma$, where $\sigma$ is the standard 
deviation of the best fit Gaussian function. With this method, we find that $R=$0.014 pc for OMC3-mms7 and 0.020 pc for 
OMC3-m. The choice of $R =2 \sigma$ is based on the consideration 
that it includes sufficient information inside the core, but excludes 
the surrounding gas along the filament around the core. 
To examine whether the choice of the core radius affects our conclusion,  
here we calculate the Collapsing Indexes with $R$ set to 1$\sigma$ and 3$\sigma$. 
The resulting CI values are listed in Table \ref{tab:diffci}, which shows that, for
both OMC3-mms7 and OMC3-m, the different choices for the core radius 
change the CI only slightly. This suggests that the determination of the cores' dynamic status by the CI
is insensitive to the choice of the core radius. Therefore, our conclusion concerning the dynamic status of 
the cores selected in this study appears to be robust.

\begin{table}
\bc
\small
\caption{Collapsing Index when choosing different core radii.} \label{tab:diffci}
\small
\begin{center}
\begin{tabular}{cccc}
\hline
\hline
				  & CI (R=1$\sigma$) & CI (R=2$\sigma) ^{*}$	& CI (R=3$\sigma)$ \\ \hline
 OMC3-mms7 & 1.26   &   1.90   & 1.39 \\ \hline
 OMC3-m &	0.06 &     0.32  &  0.39	 \\ \hline
\end{tabular}
\end{center}
\ec
\tablecomments{0.86\textwidth}{Here $\sigma$ is the standard deviation of the best fit Gaussian function to the intensity map. 
For OMC3-mms7, $\sigma$ is 0.007 pc and for OMC3-m, $\sigma$ is 0.01 pc. 
$^ {*}\,$These radii are used in the main text.}
\end{table}

\section{the $^{13}$CO and C$^{18}$O line profiles of OMC3-mms7 and OMC3-m}{\label{co}}

\begin{figure*}[ht!]
 \begin{center}
           \includegraphics[angle = 0, trim =0cm 0cm 0cm 0cm,width=12cm]{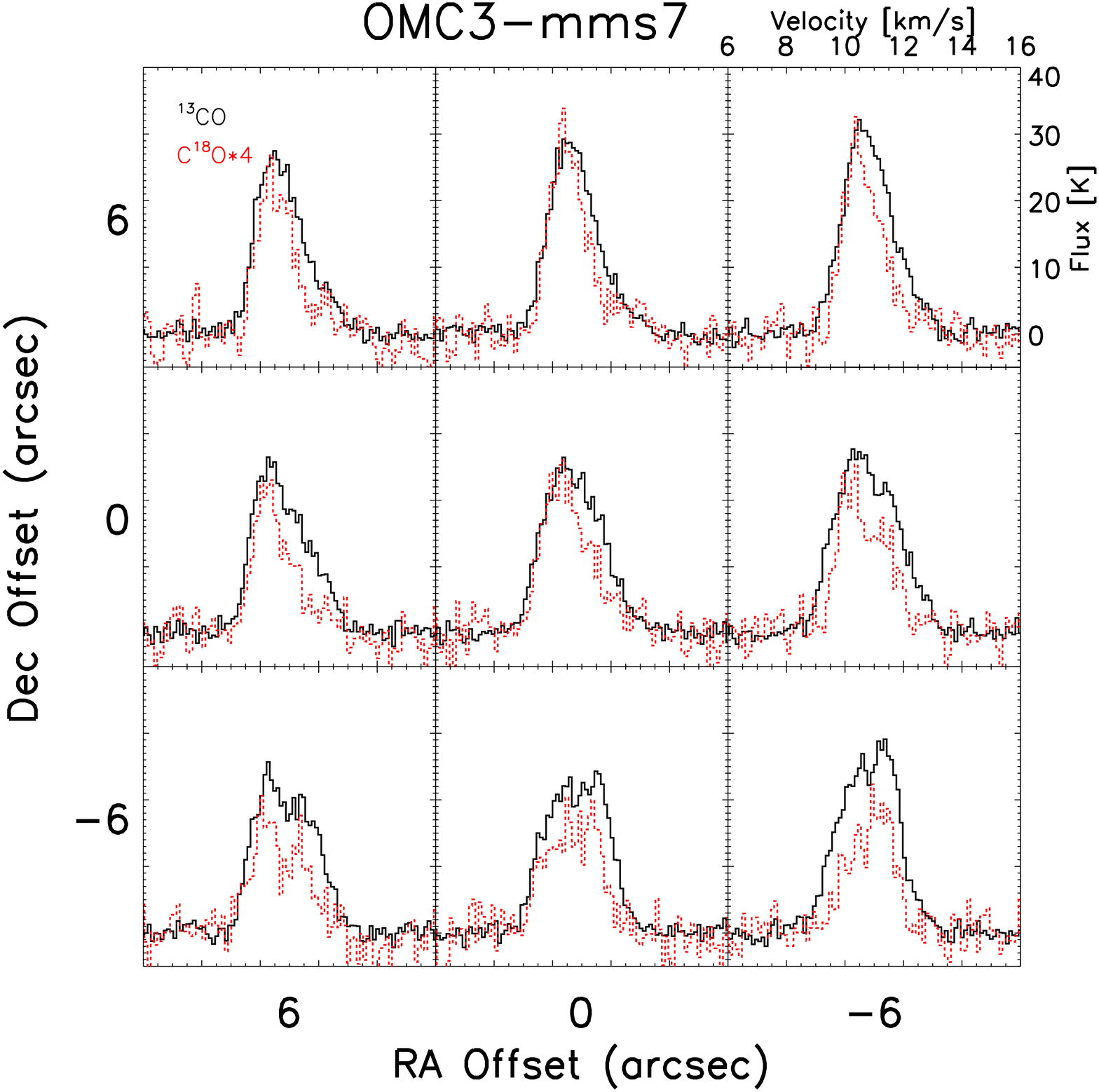}
       \end{center}
	\caption{ The $^{13}$CO J = 1-0 (black solid lines) and C$^{18}$O J = 1-0 (red dashed lines) molecular lines map in the OMC3-mms7 core. 
	The spectra are taken from the CARMA-NRO Orion Survey with a resolution of around 8" \citep{kong2018}. The whole map is centered 
	around OMC3-mms7 (R. A. = $5^h 35^m 26^s.4$, DEC. =$-5^\circ 4'1''.2$) with a grid spacing of 6", and covers a region 
	size of 18" $\times$ 18". The radius of OMC3-mms7 is 7.2" (0.014 pc).}
	\label{fig:line_omc3_mms7}
\end{figure*}

\begin{figure*}[ht!]
 \begin{center}
           \includegraphics[angle = 0, trim =0cm 0cm 0cm 0cm,width=12cm]{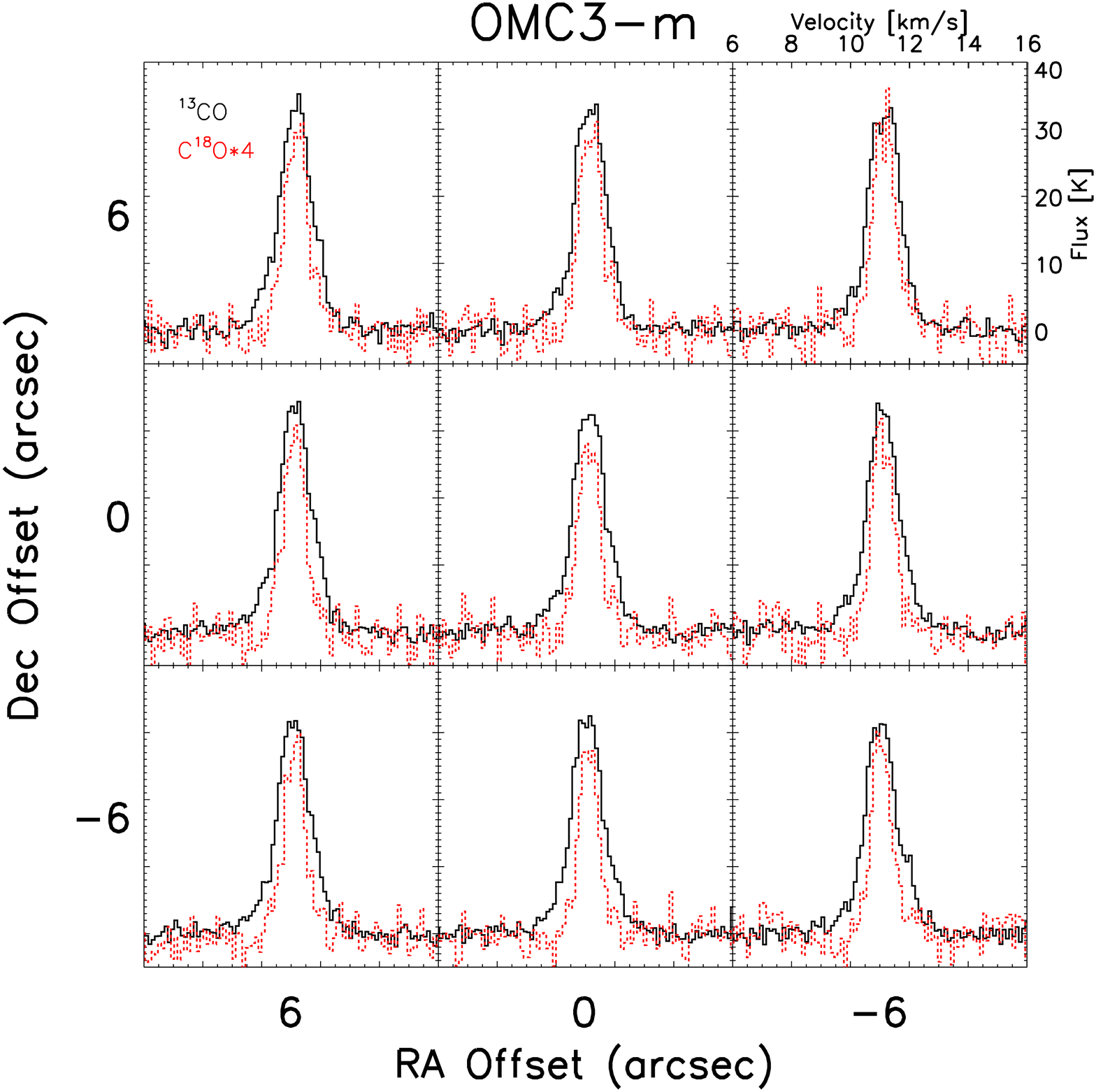}
       \end{center}
	\caption{Maps of the $^{13}$CO J = 1-0 (black solid lines) and C$^{18}$O J = 1-0 (red dashed lines) molecular lines in the OMC3-m core. The whole map is centered around 
	OMC3-m  (R. A. = $5^h 35^m 25^s.3$, DEC. =$-5^\circ 2'36''.7$) with a grid spacing of 6", and covers a region size of 18" $\times$ 18". The radius of OMC3-m is 10" (0.02 pc).}
	\label{fig:line_omc3_m}
\end{figure*}

In Fig.\ \ref{fig:line_omc3_mms7} and Fig.\ \ref{fig:line_omc3_m}, we show the $^{13}$CO (J = 1-0) and C$^{18}$O (J = 1-0)  line profiles in OMC3-mms7 and OMC3-m, 
respectively. The datacubes are taken from the CARMA-NRO Orion Survey \citep{kong2018}. The observations were carried by the Combined Array of Research in Millimeter-wave 
Astronomy (CARMA) between 2013 and 2015. Combined with the single-dish data from the Nobeyama telescope, this survey provides a map of the Orion A cloud with high spatial 
resolution and high dynamical range. The beam sizes  are 8" $\times$ 6" for the observation of $^{13}$CO and 10" $\times$ 8" for C$^{18}$O. 
The rms noises of $^{13}$CO and C$^{18}$O maps are 0.64 K and 0.47 K per channel, respectively. They have the same velocity 
resolution of 0.22 km/s.
For OMC3-m (Fig.\ \ref{fig:line_omc3_m}), the lines are rather symmetric, indicating no significant infall motion. 
On the other hand, the lines in OMC3-mms7 (Fig.\ \ref{fig:line_omc3_mms7}) appear to be asymmetric, 
and the stronger blue-shifted shoulders likely show signatures of infall. 
These are consistent with our conclusion in the text, supporting the validity of our new approach to detect the dynamical status of prestellar cores 
based on the CI measurement.

\end{document}